\documentclass{article}

\input{epsf}

\newcommand{\mathbb}[1]{{\bf{#1}}}
\newcommand{\text}[1]{{{#1}}}

\newcommand{\notag}{\nonumber}

\begin{document}

\begin{center}
{\LARGE Some Computations with Seiberg--Witten Invariant Actions\medskip }

Lorenzo Cornalba\smallskip

Laboratoire de Physique Th\'{e}orique de l'\'{E}cole Normale Sup\'{e}rieure

Paris, France\medskip
\end{center}

\textit{We show, with a }$2$\textit{--dimensional example, that the low
energy effective action which describes the physics of a single }$D$\textit{%
--brane is compatible with }$T$\textit{--duality whenever the corresponding }%
$U\left( N\right) $\textit{\ non--abelian action is form--invariant under
the non--commutative Seiberg--Witten transformations.}
\footnote{Contributions to the conference {\it BRANE NEW WORLD and
Noncommutative Geometry}, Torino, Villa Gualino, (Italy) October, 2000.}

\section{\label{secIntro}Introduction and Basic Setting}

It has been shown, by Seiberg and Witten \cite{SW}, that, in the presence of
a background magnetic field, the dynamics on $D$--branes can be described in
terms gauge fields on a non--commutative space. Most of the subsequent
literature has been devoted to the study of the low--energy $\alpha ^{\prime
}\rightarrow 0$ limit, which yields a non--commutative Yang--Mills quantum
field theory. In \cite{C} the author has considered, on the other hand, the
full effective action, to all orders in $\alpha ^{\prime }$, which is
considered as a \textit{classical} action which describes, at tree level,
the scattering of gluons at weak coupling but arbitrarily high energies. As
described in the work of Seiberg and Witten, in the presence of a background 
$B_{ab}$ field, one can describe the physics on the brane either using
ordinary gauge fields or, after an appropriate field redefinition, using
non--commutative gauge fields. The form of the action, on the other hand, is
the same in the two descriptions. This fact is quite non--trivial, and
highly constrains the form of the non--abelian Born--Infeld action, together
with all the derivative corrections.

In particular, in \cite{C}, the author has shown how to construct explicitly
to all orders in $\alpha ^{\prime }$, invariant non--abelian actions in $2$
dimensions. We show in this note how these actions, in the special $U\left(
1\right) $ abelian case, are compatible with $T$--duality. This shows that
Seiberg--Witten invariance is the correct framework to use in the analysis
of the physics on brane world--volumes. In particular, this paper represents
a partial step towards an explicitly covariant description of the brane
world--volume action, whose invariance group in enhanced from the usual
geometrical invariance of diffeomorphisms on the world--volume to a more
exotic invariance, of which we still lack a clear geometric understanding.

Let us then consider a flat space time, and let us look at the physics of $N$
flat $D$--branes with a $2$--dimensional world volume $M$. We let $x$ and $y$
be the coordinates on the world--volume $M$, and we will concentrate only on
the physics in the $x$--$y$ plane, neglecting the dynamics in the transverse
directions in space--time. Finally, we will use throughout a metric $\delta
_{ab}$ on $M$ with \textit{Euclidean} signature and we will use units such
that $2\pi \alpha ^{\prime }=1$.

The physics on the brane world--volume, at weak string coupling, is
described by an effective action written in terms of a $U\left( N\right) $
gauge--potential $A_{a}$ (with corresponding field--strength $F_{ab}$),
which is given as an $\alpha ^{\prime }$ expansion 
\begin{equation}
S=\int d^{2}x\,\mathrm{Tr}\left( 1+\frac{1}{4}F^{2}+\cdots \right) .
\label{eq200}
\end{equation}

Let us, for the moment, consider the case $N=1$, and let us perform a $T$%
--duality in the $y$ direction. Following basic facts on $T$--duality \cite
{WATI}, we are now considering the physics of a $D$--brane with $1$%
--dimensional world--volume, moving in the $x$--$y$ plane. The action then
depends on the motion of a point on the Euclidean plane $M$ -- \textit{i.e.}
a curve $\Gamma $ in $M$. More precisely, following \cite{WATI}, one must
consider the action $S$, dimensionally reduced in the $y$ direction, where
we may consider $A_{x}=0$ and where 
$$
A_{y}\left( x\right) \rightarrow Y\left( x\right)
$$
represents a specific parametrization of the curve $\Gamma $ in the plane $M$%
. Let us call $\widetilde{S}$ the $N=1$ dimensionally reduced action, coming
from $S$.

When we move from $S$ to $\widetilde{S}$, we obtain an action written in
terms of a specific parametrization $Y\left( x\right) $ of the path $\Gamma $
in the plane $M$. On the other hand, from a purely geometric point of view,
the action must be invariant under reparametrizations of $\Gamma $. In
particular, given the simplicity of the geometry, it is easy to describe all
possible invariant actions $\widetilde{S}$. First of all, let us define the
velocity and acceleration 
$$
V=\dot{Y}\text{ \ \ \ \ \ \ \ \ \ \ \ \ \ \ }A=\ddot{Y},
$$
where the dot represents differentiation with respect to $x$. The invariant
length $dl$ on the path $\Gamma $ and the extrinsic curvature $R$ are
respectively given by 
$$
dl=\left( 1+V^{2}\right) ^{%
%TCIMACRO{\UNICODE[m]{0xbd}}%
%BeginExpansion
{\frac12}%
%EndExpansion
}\text{\thinspace }dx\text{ \ \ \ \ \ \ \ \ \ \ \ \ \ \ \ \ \ \ \ \ \ \ \ \
\ \ }R=\frac{A^{2}}{\left( 1+V^{2}\right) ^{3}}.
$$
Finally, covariant differentiation with respect to arc--length is given by 
$$
\frac{d}{dl}=\left( 1+V^{2}\right) ^{-%
%TCIMACRO{\UNICODE[m]{0xbd}}%
%BeginExpansion
{\frac12}%
%EndExpansion
}\frac{d}{dx}.
$$
The possible terms which appear in the action are then integrals, with
respect to arc--length $dl$, of powers of $R$ and of its covariant
derivatives $\frac{d^{n}R}{dl^{n}}$. Examples of possible terms are 
\begin{equation}
\int dl\text{ \ \ \ \ \ \ \ \ \ \ \ \ \ \ }\int dl\,R^{n}\,\ \ \ \ \ \ \ \ \
\ \ \ \ \ \int dl\,R^{n}\frac{dR}{dl}\frac{dR}{dl}.  \label{eq100}
\end{equation}

The description of the possible structures of $\widetilde{S}$ just given is
extremely natural (and in this setting almost trivial). On the other hand,
from the point of view of the original $2$--dimensional action $S$, written
in terms of the $U\left( 1\right) $ gauge potential $A_{a}$, the fact that
one obtains, after $T$--duality, actions written in terms of monomials like (%
\ref{eq100}), is a quite remarkable numerical coincidence. Let us, for
instance, consider the first term in (\ref{eq100}). As it is well known,
this corresponds, from the $2$--dimensional point of view, to the abelian
Born--Infeld action. We will denote, in what follows, with $F\equiv F_{xy}$
the unique component of the field--strength. Under $T$--duality this
corresponds to the velocity 
$$
F\rightarrow_T \,\,V.
$$
Therefore, the length of the path $\Gamma $ is given by 
$$
\int dl=\int dx\left( 1+V^{2}\right) ^{%
%TCIMACRO{\UNICODE[m]{0xbd}}%
%BeginExpansion
{\frac12}%
%EndExpansion
}\rightarrow \int d^{2}x\left( 1+F^{2}\right) ^{%
%TCIMACRO{\UNICODE[m]{0xbd}}%
%BeginExpansion
{\frac12}%
%EndExpansion
}=\int d^{2}x\,\left( 1+\frac{1}{2}F^{2}-\frac{1}{8}F^{4}+\cdots \right) .
$$
The fact that the LHS of the above equality has a geometric interpretation
imposes restrictions on the $\alpha ^{\prime }$ expansion coefficients which
appear on the RHS of the equality. More generally, the $\alpha ^{\prime }$
expansion of the low--energy effective action $S$ is constrained by
compatibility with the geometric interpretation of $T$--duality.

Quite independently from the above considerations, the action $S$ possesses 
\cite{SW}, in the full non--abelian setting, an other interesting invariance
property. If one considers adding a central term to the field--strength $%
F_{ab}\rightarrow F_{ab}+\mathbf{1}\cdot B_{ab}$, this has the same effect
as rewriting the same action on a new world--volume $\widetilde{M}$, which
is a non--commutative deformation of the original Euclidean plane $M$, given
by the relation $\left[ x,y\right] =i\theta ^{xy}$. The tensor $\theta $ is
given by $\theta =B\left( 1+B^{2}\right) ^{-1}$. Moreover, the metric on the
new plane $\widetilde{M}$ is given by $\left( 1+B^{2}\right) $. The new
action is now written in terms of a non--commutative Yang--Mills potential
(explicitly given in terms of the original $A_{a}$ and $\theta ^{ab}$) and
has the \textit{same structure} as $S$. This fact imposes severe constraints
on the original $\alpha ^{\prime }$ expansion given in (\ref{eq200}), as was
analyzed in detail in \cite{C}, and in particular relates terms of order $%
\left( \alpha ^{\prime }\right) ^{L}$ to those of order $\left( \alpha
^{\prime }\right) ^{L-1}$ and $\left( \alpha ^{\prime }\right) ^{L+1}$. We
will call the above restriction Seiberg--Witten (SW) invariance of $S$.

We will show in this note that the SW invariance of $S$ in the non--abelian
setting implies, in a quite non--trivial way, consistency under $T$--duality
of the $U\left( 1\right) $ action, as discussed above.

\section{Invariant Actions}

In this section we review how to construct SW\ invariant action in $2$%
--dimensions. We will describe the algorithm, but we will not try to
motivate it, since this would imply a long digression. The interested reader
can consult \cite{C}, where the full SW invariance is discussed in detail.

First we introduce on $M$ complex coordinates $z,\overline{z}=\frac{1}{\sqrt{%
2}}\left( x\pm iy\right) $. Corresponding to these coordinates, we introduce 
\textit{formal letters }$Z,\overline{Z}$, and we consider the vector space $%
W_{L}$ of \textit{cyclic} \textit{words} built with $L$ letters $Z$ and $L$
letters $\overline{Z}$. Vectors in $W_{L}$ are then linear combinations of
strings of $2L$ letters, where we consider two strings equal if they differ
only by a cyclic permutation of the letters. We will call the integer $L$
the level, and this will correspond, as we will see below, to the level in
the $\alpha ^{\prime }$ expansion (with the $F^{2}$ term at level $2$).

We will actually be more interested in the subspace $G_{L}\subset W_{L}$ of
so--called \textit{gauge invariant words}, which are defined as follows. Let 
$a$,\thinspace $a^{\dagger }$ be standard creation and annihilation
operators satisfying $\left[ a,a^{\dagger }\right] =1$ and let $O$ be the
space of cyclic polynomials in $a,a^{\dagger }$. If we define the map $%
r:W_{L}\rightarrow O$ by $Z,\overline{Z} \mapsto a,a^{\dagger }$%
, then $G_{L}=\ker r$. Elements in $G_{L}$ correspond to terms in $S$ at
level $\left( \alpha ^{\prime }\right) ^{L}$. More precisely, we have the
correspondences 
\begin{eqnarray}
F &\rightarrow &Z\overline{Z}-\overline{Z}Z  \label{eq300} \\
D_{z}\cdots  &\rightarrow &i\left[ Z,\cdots \right] \text{ \ \ \ \ \ \ \ \ \
\ \ \ \ \ \ \ \ \ \ }D_{\overline{z}}\cdots \rightarrow i\left[ \overline{Z}%
,\cdots \right]   \notag
\end{eqnarray}
where $D_{a}$ is the $U\left( N\right) $ covariant derivative. For example,
one can check that $\dim G_{2}=1$, and that the unique element is given by $Z%
\overline{Z}Z\overline{Z}-ZZ\overline{Z}\overline{Z}$, which, following the
above identifications, corresponds to the term $\frac{1}{2}\int d^{2}x\,%
\mathrm{Tr}F^{2}$ at level $2$. It is also clear that, given any expression
in $S$, one can use the identifications (\ref{eq300}) to rewrite it in terms
of words in $Z,\overline{Z}$. The words will automatically be cyclic, since
the action $S$ is given by taking the ``trace'' $\int d^{2}x\,\mathrm{Tr}$
of a local expression in the field strength. Moreover, the original gauge
invariance of the action is reflected in the fact that the resulting words
are elements of the subspace $G_{L}$.

To construct SW--invariant actions we need to introduce two operators --
denoted $\Delta ,\overline{\Delta }$ -- which act naturally on the spaces $%
G_{L}$ as 
$$
\Delta :G_{L}\rightarrow G_{L+1}\,\ \ \ \ \ \ \ \ \ \ \ \ \ \ \ \ \ \ \ \ \
\ \overline{\Delta }:G_{L}\rightarrow G_{L-1}.
$$
In particular, the operator $\Delta $ is defined as a derivation which acts
on the single letters as 
$$
\Delta Z=\frac{1}{4}(ZF+FZ)\,\ \ \ \ \ \ \ \ \ \ \ \ \ \ \ \ \ \ \ \ \ \ \
\Delta \overline{Z}=\frac{1}{4}(\overline{Z}F+F\overline{Z}).
$$
Similarly $\overline{\Delta }$ is defined by 
$$
\overline{\Delta }F=1.
$$
More precisely, given a word $w\in G_{L}$, we write it in terms of the
field--strength $F$ (and of its covariant derivatives) using the
correspondence (\ref{eq300}). We then apply $\overline{\Delta }$ on each $F$
as a derivation. It is quite a non--trivial fact \cite{C} that$\,[\overline{%
\Delta },\Delta ]=2L-1$.

With this notation in place, we can now easily describe SW--invariant
actions, following the results of \cite{C}. A general invariant action is
given by arbitrary linear combinations of what we will call \textit{%
invariant blocks}. An invariant block is itself constructed starting from a 
\textit{lowest level term} $g_{P}\in G_{P}$, for some integer $P$, and
contains terms in the $\alpha ^{\prime }$ expansion of levels $L\geq P$,
with $L-P\in 2\mathbb{N}$. The term $g_{P}$ must satisfy the basic equation 
$$
\overline{\Delta }g_{P}=0,
$$
and the invariant block is then constructed as follows. We construct terms $%
g_{L}\in G_{L}$ for $L>P$ as 
$$
g_{L+1}=\Delta g_{L}.
$$
It follows from $\left[ \overline{\Delta },\Delta \right] =2L-1$ that $%
\overline{\Delta }g_{L+1}=c_{L,P}g_{L}$, where 
$$
c_{L,P}=(L+P-1)(L-P+1).
$$
Construct then coefficients $d_{L}$ for $L-P\in 2\mathbb{N}$ defined by $%
d_{P}=1$ and by the recursion relation 
$$
d_{L+2}=-\frac{d_{L}}{c_{L+1,P}}.
$$
The invariant block constructed from $g_{P}$ is then given by 
$$
\sum_{L-P\in 2\mathbb{N}}d_{L}\,g_{L}.
$$

\section{Abelian Actions}

Up to now, we have discussed how to construct general $U\left( N\right) $
actions which are SW invariant. We now wish to analyze in more detail the
invariant actions in the abelian $N=1$ case. As is well known, in this case
one can classify terms in the action $S$ not only by the level $L$ in the $%
\alpha ^{\prime }$ expansion, but also by the number $D$ of derivatives
appearing in the various terms. For example, a term like $F^{2}\partial
F\partial F$ has $L=5$ and $D=2$. It is not difficult to show from the
definitions (see \cite{C}) that the operation $\Delta $ increases $L$ by $1$
and \textit{does not decrease} $D$. On the other hand, the operator $%
\overline{\Delta }$ decreases $L$ by one unit and \textit{leaves }$D$ 
\textit{unchanged}.

Let us now consider an invariant block with a lowest term $g_{P}$ with level 
$P$ and number of derivatives $D$. We claim that, with very little
computation, we can resum all the terms in the invariant block with exactly $%
D$ derivatives. To show this, first recall that $\overline{\Delta }%
^{L-P}g_{L}\propto g_{P}$. Since $g_{P}$ contains $D$ derivatives, given the
properties of $\Delta ,\overline{\Delta }$ one has 
$$
\left. \overline{\Delta }^{L-P}g_{L}\right| _{D\text{\textrm{derivatives}}%
}\propto g_{P}.
$$
Moreover, since the action of $\overline{\Delta }$ on terms with $D$
derivatives is completely known, one can quickly reconstruct the full
invariant block. We will give below a concrete example, which will clarify
the general discussion.

We will show that the $D$ derivative part of an invariant block is
explicitly given by the covariant terms described in section \ref{secIntro}
(as in equation (\ref{eq100})). A generic covariant term is constructed with 
$N_{R}$ powers of curvature $R$ and with $N_{D}$ covariant derivatives $%
\frac{d}{dl}$. Recalling the correspondences $V\rightarrow F$, $A\rightarrow
\partial F$, $\cdots $ it is simple to show that 
$$
P=3N_{R}+\frac{1}{2}N_{D}\,\ \ \ \ \ \ \ \ \ \ \ \ \ \ \ \ \ \ \ \ \ \ \ \
D=2N_{R}+N_{D}.
$$
Since $N_{R},N_{D}\geq 0$, we then have that 
$$
\frac{1}{2}D\leq P\leq \frac{3}{2}D.
$$
In Figure $1$ we show a graph of the possible $P,D$ combinations for lowest
level terms, and we show the corresponding covariant terms.

\epsfbox{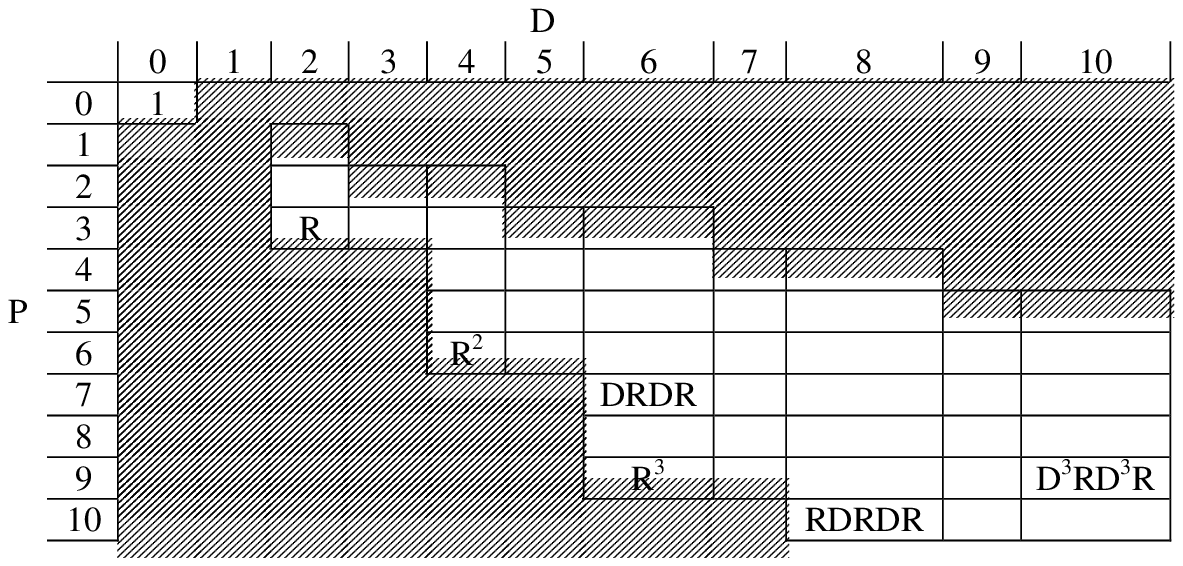}

{\bf Figure 1.} Derivative with respect to arc--length $dl$ is denoted by
$D$ in the figure.

\section{Examples}

Let us consider the covariant term 
\begin{equation}
\int dl\,R^{n}=\int dx\frac{A^{2n}}{\left( 1+V^{2}\right) ^{3n-%
%TCIMACRO{\UNICODE[m]{0xbd}}%
%BeginExpansion
{\frac12}%
%EndExpansion
}}  \label{eq400}
\end{equation}
We clearly have that $P=3n$ and $D=2n$. Explicitly one has (we are using the 
$T$--duality correspondences $F\rightarrow V$, $\cdots $) 
\begin{eqnarray*}
g_{P} &=&A^{2n}\,\ \ \ \ \ \ \ \ \ \ \ \ \ \ \ \ \ \ \ \ \ \ \ \ \ \ \ \ \ \
\ \ \ \ \ \ \ \ \ \ g_{P+1}=\left( 2P-1\right) VA^{2n}+\cdots \\
g_{P+2} &=&2P\left( 2P-1\right) V^{2}A^{2n}+\cdots
\end{eqnarray*}
and $d_{P}=1$, $d_{P+2}=-\frac{1}{4P}$. Therefore the action starts as

$$
\int dx\,A^{2n}\left[ 1+\left( -3n+\frac{1}{2}\right) V^{2}+\cdots \right]
$$
thus reproducing the first terms in (\ref{eq400}), and in particular the
non--trivial coefficient $\left( -3n+\frac{1}{2}\right) $. In fact we can
show easily that we obtain all the terms in (\ref{eq400}). Generally we have
that $g_{L}=k_{L}V^{L-P}A^{2n}$ where the coefficients $k_{L}$ satisfy $%
k_{P}=1$ and $\left( L+1-P\right) k_{L+1}=c_{L,P}k_{L}$, or $k_{L+1}=\left(
L+P-1\right) k_{L}$. The recursion relation is satisfied by 
$$
k_{L}=\frac{\left( L+P-2\right) !}{\left( 2P-2\right) !}.
$$
Moreover the coefficients $d_{L}$ are given by 
$$
d_{P+2q}=\left( -\right) ^{q}\frac{1}{2^{2q}}\frac{\left( P-1\right) !}{%
\left( P+q-1\right) !q!}.
$$
Putting all together we see that the action is then given by 
\begin{eqnarray*}
\sum_{q=0}^{\infty }d_{P+2q}g_{P+2q} &=&\sum_{q=0}^{\infty }\left( -\right)
^{q}\frac{1}{2^{2q}}\frac{\left( P-1\right) !}{\left( P+q-1\right) !q!}\frac{%
\left( 2P+2q-2\right) !}{\left( 2P-2\right) !}\int dx\,V^{2q}A^{2n} \\
&=&\int dx\frac{A^{2n}}{\left( 1+V^{2}\right) ^{3n-%
%TCIMACRO{\UNICODE[m]{0xbd}}%
%BeginExpansion
{\frac12}%
%EndExpansion
}}.
\end{eqnarray*}

We have given here an example of the consistency of the construction of SW
invariant actions. Many more examples can be given, all with the same basic
conclusion, that SW invariance implies consistency under $T$--duality in the
abelian case. We leave a more complete discussion to \cite{C2}.

\end{document}